
\input harvmac

\baselineskip=20pt

\def\ni{\noindent}
\def\a{\alpha}
\def\b{\beta}
\def\g{\gamma}
\def\h{\eta}
\def\i{\iota}
\def\l{\lambda}
\def\L{\Lambda}

\def\s{\sigma}
\def\bi{\bar \i}

\def\gh{{\bf \hat g}} \def\gg{{\bf g}} 

\def\EM{\hat E_M}
\def\en{\hat E_N}
\def\e1{\hat{E}_1 }
\def\em1{\hat{E}_{-1} }
\def\fiz{\hat F^i(z)}

\def\fibz{\hat F^{\bar \i}({\bar z)}}

\def\fih{\hat F^i(\eta)}

\def\rlj{|\Lambda_j>}\def\rlk{|\Lambda_k>}
\def\llj{<\Lambda_j|}\def\llk{<\Lambda_k|}
\def\rlo{|\Lambda_0>}
\def\llo{<\Lambda_0|}

\def\rlj{|\Lambda_j>}\def\rlk{|\Lambda_k>}
\def\llj{<\Lambda_j|}\def\llk{<\Lambda_k|}
\def\rlo{|\Lambda_0>}
\def\llo{<\Lambda_0|}

\Title{ Swansea SWAT/92-93/6}
{\vbox{\centerline{Crossing and Antisolitons in}
\centerline{Affine Toda Theories}}}


\centerline{Marco A.C. Kneipp and David I. Olive}

\centerline{Department of Mathematics,
University College of Swansea,}
\centerline{ Swansea SA2 8PP, Wales, UK.}

\centerline{\bf Abstract}

\baselineskip=12pt

Affine Toda theory is a relativistic integrable theory in two dimensions
possessing solutions describing a  number of  different species of solitons
when the coupling is chosen to be imaginary. These nevertheless carry real
energy and momentum. To each species of soliton there has
to correspond an antisoliton species. There are two different ways of
realising  the  antisoliton whose equivalence is shown to follow from a
surprising identity satisfied within the  underlying affine Kac-Moody group.
This is the classical analogue of the crossing property of analytic S-matrix
theory. Since a complex parameter  related to the coordinate
of the soliton is inverted, this identity implies a sort of modular
transformation  property of the soliton solution. The results simplify
calculations of  explicit soliton solutions.

\baselineskip=24pt

\Date{May 1993} 



\bigskip
\ni{\it 1. Introduction}
\medskip

As integrable field theories in two dimensions, the affine Toda theories have
long attracted particular attention \ref\AFZ{ A.E. Arinshtein, V.A. Fateev
and A.B. Zamolodchikov, {\it Phys. Lett} {\bf B87} (1979) 389-392,
\lq\lq Quantum S-Matrix of the $1+1$ dimensional Toda Chain"},
\ref\MOP{ A.V. Mikhailov, M.A. Olshanetsky
and A.M. Perelomov, {\it Comm. Math. Phys.} {\bf 79} (1981)
473-488; \lq\lq Two-Dimensional Generalised Toda Lattice" },
\ref\LSS{ A.N. Leznov, M.V. Saveliev and V.G. Smirnov, {\it Sov. J. Theor.
Math. Phys. } {\bf 48} (1981) 3-12}, \ref\W{ G. Wilson, {\it Ergod. Th. \&
Dynam. Sys.}  {\bf 1} (1981) 361-380, \lq\lq The modified Lax and
two-dimensional Toda lattice equations associated with simple Lie algebras"},
 \ref\OTa{ D.I Olive and N. Turok,{\it Nucl. Phys.}  {\bf B215 [FS7]}
(1983) 470-494, \lq\lq The Symmetries of Dynkin Diagrams and the
Reduction of Toda Field Equations"}, \ref\OTc { D.I. Olive and N. Turok,
{\it Nucl.Phys.} {\bf B257 [FS14]}
(1985) 277-301, \lq\lq Local Conserved Densities and Zero Curvature
Conditions for Toda Lattice Field Theories"}, \ref\OTd{ D.I. Olive and N.
Turok,  {\it Nucl.Phys.} {\bf B265 [FS15]}
(1986) 469-484, \lq\lq The Toda Lattice Hierarchies and Zero
Curvature Conditions in  Kac-Moody Algebras"}. There were two main reasons:
(1) their relativistic   invariance and (2) their
natural association with affine Kac-Moody algebras. These algebras (which we
shall take to be untwisted and simple laced  for definiteness and simplicity)
play a key r\^ole in the integrability which is manifested by the occurrence
of an infinite number of local conservation laws. Correspondingly, there
exists a hierarchy of integrable systems amongst which the affine Toda theory
is distinguished by its Poincar\' e invariance.

This  significance was reinforced by Zamolodchikov' observation that the
Lorentz symmetry is a subgroup of a much large, conformal symmetry realised
in a critical limit of the theory  \ref\ZamB{ A.B. Zamolodchikov,
{\it  Int. J. Mod. Phys.} {\bf A4} (1989) 4235-4248, \lq\lq Integrals of
Motion and S-Matrix of the (Scaled) $T=T_c$ Ising Model with Magnetic
Field"},  \ref\ZamC{ A.B. Zamolodchikov , \lq\lq Integrable Field Theory from
Conformal Field Theory", in \lq\lq Integrable Systems in Quantum Field Theory
and
Statistical  Mechanics", {\it Advanced Studies in Pure Mathematics"} {\bf 19}
(1989) 641-674 (Academic Press)}. The conservation laws are seen as the
non-chiral relics  of the conformal (or larger)
chiral symmetry when it is broken to its Poincar\' e subgroup.
Affine Toda theory provides a very good
illustration  of this \ref\HMb{T. Hollowood and P. Mansfield, {\it Phys.
Lett.} {\bf 226B} (1989) 73-79, \lq\lq Rational Conformal Field Theories at,
and away from, Criticality as Toda Field Theories"}.

When  the affine Toda couplings are made imaginary, the theory acquires
classical soliton solutions \ref\Hol92{ T.J. Hollowood, {\it Nucl. Phys.}
{\bf B384} (1992) 523-540, \lq\lq Solitons in Affine Toda Field Theories"},
\ref\MM{ N.J. MacKay and W.A. McGhee;  preprint DTP-92-45/RIMS-890 ; \lq\lq
Affine Toda Solitons and Automorphisms of Dynkin Diagrams"},  \ref\Ar{
H.Aratyn, C.P. Constantinidis, L.A. Ferreira, J.F. Gomes and A.H. Zimerman,
\lq\lq Hirota's Solitons in the Affine and the Conformal Affine Toda Models",
IFT-P.052/92 preprint} whose energy and momentum are topological in the sense
of being surface terms and, as a consequence, real despite the complex nature
of the equations
 \ref\LOT{ H.C. Liao, D.I. Olive and N. Turok, {\it Phys.
Lett.} {\bf B298} (1993) 95-102, \lq\lq Topological Solitons in $A_r$
Affine Toda Theory" }, \ref\OTUa{ D.I. Olive, N. Turok and
J.W.R.Underwood, Solitons and the Energy-Momentum Tensor for Affine Toda
Theory; Preprint Imperial/TP/91-92/35 and SWAT/3, to appear in {\it Nucl.
Phys.}},  \ref\Nie{ M.R. Niedermaier, The Spectrum of the Conserved Charges
in Affine Toda Theories, DESY preprint 92-105}. If $ \gh $
denotes  the associated untwisted affine Kac-Moody algebra, with the Lie
algebra $ \gg $ simple and of rank $ r $, the number of soliton species
equals $r$, \OTUa, the same as the number of species of particles created by
the quantum Toda fields. Furthermore, there are resemblances  between the
solitons and quantum excitation particles such as the mass formulae and the
rules governing trilinear couplings \ref\BCDa{ H.W. Braden, E. Corrigan, P.E.
Dorey and R. Sasaki,  {\it Phys. Lett.} {\bf B227} (1989) 411: \lq\lq
Extended Toda Field  Theory and Exact S-Matrices"},
\ref\BCDSb{ H.W. Braden, E. Corrigan, P.E. Dorey and R. Sasaki,
{\it Nucl. Phys.} {\bf B338} (1990) 689:
\lq\lq Affine Toda field Theory and Exact S-Matrices"},
 \ref\KM{ T.R. Klassen and E. Melzer, {\it Nucl. Phys.} {\bf B338} (1990)
485-528, \lq\lq Purely Elastic Scattering theories and Their Ultraviolet
Limits"}, \ref\CM{ P. Christe and G. Mussardo, {\it Int. J. Mod Phys.} {\bf
A5} (1990) 4581, \lq\lq Elastic S-matrices in $1+1$ Dimensions and Toda Field
Theories"},  \ref\Dora{ P.E. Dorey, {\it Nucl. Phys.} {\bf B358} (1991) 654,
\lq\lq Root  Systems and Purely Elastic S-Matrices"}, \ref\Fre{M.D. Freeman,
{\it Phys. Lett.}  {\bf B261} (1991) 57 ; \lq\lq On the Mass Spectrum of
Affine Toda Field Theory"}, \ref\FLO{A. Fring,
H.C. Liao and D.I. Olive, {\it Phys. Lett.} {\bf B 266} (1991)
82-86, \lq\lq The Mass Spectrum and Coupling in Affine Toda
Theories"}.

There  is an extra discrete symmetry of  a relativistic theory which implies
that to every  species  $ i $ of particles there exist a corresponding
species $\bar \i $ of antiparticles to $ i$. For the  particles created by
the quantum Toda fields there is a very specific construction of  $\bar \i $
from $i$ \FLO. But for solitons there is a second, apparently different
construction, in which the soliton solution is run backwards in  space. The
main result here is to find and demonstrate a surprising identity in terms of
the generators of $ \gh $ which guarantees the equivalence of these two
presentations of the antisoliton. For reasons to be explained, this identity
can be thought  as the classical analogue of the crossing property familiar
in analytic S-matrix theory \ref\ELOP{R.J. Eden, P.V. Landshoff, D.I. Olive
and J.C. Polkinghorne,  \lq\lq The Analytic S-Matrix", Cambridge University
Press (1965)}. The identity explains  various regularities in the explicit
soliton solutions and can simplify their calculations.

In  section 2 we  recap the soliton  formalism and how it is advantageous to
reformulate $\gh$ in
terms  of a new basis consisting of its principal Heisenberg subalgebra and
the remaining generators which are the Laurent coefficients of the $ r$
\lq\lq fields" $ \hat{F} ^1(z),  \hat{F} ^2(z), \ldots , \hat{F }^r(z) $
which ad-diagonalise the Heisenberg subalgebra and plays the r\^ole of
creating the $r$ species of solitons \OTUa.

It  has been shown that in a highest weight representation of $\gh $ at level
$x$
the powers of $\fiz $ higher than $x$ vanish \ref\OTUb{ D.I.Olive, N. Turok
and J.W.R.Underwood, Affine Toda Solitons and Vertex Operators; Preprint
IC/92-93/29, Princeton PUP-PH-93/1392 , Swansea SWAT/92-93/5}. In  section 3
we show that the highest non-vanishing power, $(\fiz)^x $, takes the form of
a vertex operator obtained by exponentiating the Heisenberg subalgebra. This
generalises a result previously known at level 1. In section 4, this vertex
operator is used to relate the power $k (\leq x) $ of $\fiz $ to the power
$(x-k) $ of  $\fibz $ where the variable $\bar z$ is obtained from $z$ by the
standard crossing procedure in a sense to be explained. This is the
surprising identity and constitutes our main result. Since it can be put in a
form involving exponentials of Kac-Moody generators (including the central
extension) it can be viewed as a group theoretic relation.

Since  the single soliton solution is expressed in terms of expectation
values of  powers of  $\fiz$ with respect to highest weight states, the ideas
mentioned relate powers $k$ and $x-k$ in a very simple way. Since powers up
to the integer part of $(x+1)/2$ only need to be considered, rather than $x$,
calculations are consequently simplified. \bigskip

\bigskip
\ni{\it 2. Preliminaries}
\medskip

Corresponding to the $(r+1)$ simple roots of the affine untwisted  Kac-Moody
algebra  $\gh $, we denote the $3(r+1)$ Chevalley generators as $e_i , f_i $
and $h_i$   $(i = 0, 1, ..., r) $ in the usual notation \ref\Kac{V.G.  Kac,
\lq\lq Infinite-dimensional Lie algebras" , Third Edition, {\it Cambridge
University Press}, 1990}, \OTUa . In the principal grade, which counts the
height of the roots by the operator

$$d' = -hL_0 + T_0^3, \eqno(2.1)$$

\ni these have grades $1, -1$ and $0$ respectively. $h $ is the Coxeter
number of $\gg$, $T^3_0 $ the intersection of its principal so(3) subalgebra
and its Cartan subalgebra and $L_0$ the Virasoro generator.

The principal Heisenberg subalgebra of $\gh$ plays an important role in the
affine Toda theories, \OTc, just as it does in other integrable theories
considered earlier,  \ref\JM{ M. Jimbo and T. Miwa, {\it RIMS, Kyoto Univ.}
{\bf 19} (1983) 943-1001,  \lq\lq Solitons and Infinite Dimensional Lie
Algebras"} and the final chapter of  \Kac. It is infinite dimensional,
possessing commutation relations

$$ \left[ \EM ,\en \right]=Mx\delta_{M+N,0}, \eqno(2.2)$$

\ni where the suffix denotes the principal grade and only takes values
equal to the exponents of $\gg $ modulo $h$, as far as the
Heisenberg  subalgebra is concerned. $x$ denotes the integer level. The
generators which ad-diagonalise the Heisenberg subalgebra can be written as
$\hat{F} (\a, z) $ where $\a$  is a root of  $\gg$ and $z$ is a complex
number

$$\left[\EM,\hat{F} (\a , z) \right]=\a\cdot q([M])z^M \hat{F} (\a , z)
\eqno(2.3)$$

\ni $q(\nu)$ is the eigenvector of the Coxeter element $\sigma$ corresponding
to
the exponent $\nu$ of $\gg$,

$$\s (q(\nu)) = e^{{2\pi i\nu \over h}} q(\nu) . \eqno(2.4) $$

It follows, with an appropriate normalization choice such as

$$ \llo \hat{F}(\a , z) \rlo = 1 , \eqno(2.5)$$

\ni that

$$\hat{F}(\s (\a),z) = \hat{F} (\a, z e^{-{2\pi i \over h}} ) . \eqno(2.6) $$

\ni Here it is sensible to choose a standard representative of each of the r
orbits of the $hr$ roots under action of $\sigma$. A convenient choice is,
\Dora, \FLO,

$$\g _i = c(i) \a _i  , \eqno(2.7) $$

\ni where $\a _i $ is a simple root of $g$ and $c(i) = \pm1$ according as
the vertex $i$ of the
Dynkin diagram is coloured black or white. Then, the Laurent coefficient of
the $r $ quantities
$\hat{F} (\g _i, z) $ taken with the $ \EM $, span $\gh$. We  further define

$$\fih = \hat{F}\left(\g_i  , ie^{\eta} e^{-{i\pi(1+c(i)) \over 2h}}
\right), \eqno(2.8)$$

\ni since this will create a soliton of  species $i$ with rapidity $\eta
$. In fact, with this choice of $z$,
$$[\hat E_{\pm 1},\fih]=-|\g_i\cdot q(1)|e^{\pm\eta}\fih.\eqno(2.9)$$
To see this interpretation,
we consider the general soliton solution of the affine Toda theory associated
with $\gg $

$$e^{-\b \l_j \cdot \Phi } = {\llj e^{-\mu \e1 x^+} g(0) e^{-\mu \em1 x^- }
\rlj \over                                       \llo e^{-\mu \e1 x^+} g(0)
e^{-\mu \em1 x^- } \rlo^{m_j}}
   \eqno(2.10)$$

\ni where  $\rlj $, $(j = 0, 1, ..., r)$, denote the highest weight states
of the $r$ fundamental
representations of $\gh$. When $j\neq0$ they have $\gg $ weights $\l_j $ and
levels $m_j$ where ${\psi \over \psi ^2} = \sum_{j=1}^{r} m_j {\a_j \over
\a_j ^2}$. $\rlo$ has $\gg$ weight zero  and unit level and can be thought of
as the vacuum.
$x^{\pm}$ are the light-cone coordinates $(t\pm x)/\sqrt2$ and the
coupling constant $\b$ is understood to be imaginary.

The Kac-Moody group element $g(0)$ contains the soliton data and consists of
a product of factors $\exp Q_k \hat{F}^{i_k}(\eta_k)$, one for each soliton
$i_k$.  $Q_k $ is a complex number upon which the coordinate of the $k$'th
soliton depends.

The time development operators $\exp (-\mu \hat{E}_{\pm 1} x^{\pm}) $ can be
eliminated using the fact that $\en$ annihilates  $\rlj $ for $N>0$ and the
commutation relations  (2.2) and (2.3). The  result is that $g(0)$ is
replaced by a space-time dependent $g(t)$ in which each $\fih $ is multiplied
by the factor

$$W_i = \exp \left(\mu |\g_i \cdot q(1)| (x^+ e^{\eta} - x^- e^{-\eta})\right).
\eqno(2.11)$$

\ni The choice $z=ie^{\eta}exp-\{{i\pi(1+c(i))\over2h}\}$ in (2.8) ensured
that  $W_i $ is real and that
when $\eta = 0 $ the
time, $t$,  cancels out of $W_i$ so that the soliton $i$ is then stationary.
More generally, we see that $\eta$ is its rapidity. The energy and
momentum of the general solution of this type has been evaluated and
found to be appropriate to this interpretation and independent of the
complex numbers $Q_k$, \OTUa.

Now we shall consider how to formulate an antisoliton. If we replace the
rapidity  $\eta \rightarrow \eta + i \pi $ in $W_i$, (2.10),  we see that the
sign of $x$ and $t$ is reversed. This means that the solution runs backwards
in space,  reversing the boundary condition and describing the antisoliton to
species $i$. But for the particles which are the quantum excitations of the
Toda fields there is a precise way of obtaining the antispecies $\bar \i$ of
$i$.  This is given by the orbit of $\s $ containing $-\g _i$ rather than $\g
_i$. Denoting its representative element (2.6),
$\g_{\bar \i}$, we conclude that the antisoliton of rapidity $\h $ can be
equally well be created by $\hat{F}^i (\eta + i\pi)$ and $\hat{F}^{\bar \i}
(\h) $. These quantities  are certainly not equal, but we shall find a
surprising, non-linear relation
 between them that will mean that the
antisoliton solutions created by them are indeed the same after the
parameters $Q_k$ are  readjusted.

Notice that the prescription that $i \rightarrow \bar \i $ can be
achieved by an analytic
continuation $\h \rightarrow \h + i\pi$, is a familiar property of the
scattering matrices in the
quantum theory of the particle excitations of the Toda fields where
 it is known as \lq\lq crossing".
Thus it will be possible to consider our identity   as a classical
 version of \lq\lq crossing"
which can be formulated purely in terms of the affine Kac-Moody group.

Notice also that since $\pm i\mu\hat{E}_{\pm 1} $ are displacement operators
along the two light cones, their commutation relations (2.1) with the
principal grade, $d'$, show that, in this soliton formalism, $d'$ has
the physical interpretation of being the Lorentz boost.


\bigskip
\ni{\it 3. The Highest Non-Vanishing Power of $\fih $ as a Vertex Operator}
\medskip

The exponential factor  $\exp Q \fih $ responsible for creating a soliton of
species $i$ and rapidity $\eta$ can be expanded as a power series which
terminates as a consequence of the vanishing of those
powers of $\fih $ greater than the level $x$ of the representation
considered. (This statement is slightly more complicated if $\gg$
 is not simply laced). Consequently, each expression
(2.10) for the soliton solutions comprises  a ratio of polynomials
in the $W_i$, (2.11), of the same order. This has been proven, \OTUb,
starting with the known result
that, at level $x = 1$, $\fih$ can be represented as a vertex operator
involving exponentials of elements of the Heisenberg subalgebra (2.2)
\ref\KKLW{ V.G. Kac, D.A. Kazhdan, J. Lepowsky,  and R.L. Wilson, {\it
Advances in Math.}  {\bf 42} (1981) 83-102}  . We shall now extend this
result by showing that, at level $x$, the $x$'th power of $\fih$ is again
such a vertex operator. Thus the highest non-vanishing power of $\fih$ has a
particularly simple structure. This is important as it controls the
asymptotic behaviour of the soliton solutions.

We recall that in the fundamental representation  with highest weight
$\L_j$ and
level 1 (so $m_j = 1$),

$$\fih  = e^{-2\pi i \l_i \cdot \l_j } Y^i Z^i, \eqno(3.1a) $$

\ni where

$$ Y^i = \exp \sum_{N>0 } {\g_i \cdot q([N]) z^N \hat E_{-N} \over N},\qquad
      Z^i = \exp  \sum_{N>0 } -{\g_i \cdot q([N])^*  z^{-N} \en \over N} ,
\eqno(3.1b) $$

\ni and we have used the result of section 8 of \OTUb that

$$\llj \fih \rlj = e^{-2\pi i \l_i \cdot \l_j } \llo \fih \rlo  $$

\ni and the normalization choice (2.5). The phase factor
occurring here and consequently in (3.1a) has its origin in the
isomorphism between the centre of the universal covering group of
$\gg$ and $W_0$, the subgroup of the Weyl group which relates the
vacuum to the level one representations of $\gh$, \OTa.

Now let us consider another fundamental representation with weight
$\L_k$, now of level
$m_k$ greater than 1. The corresponding highest weight state $\rlk$ can be
represented as a state occuring in the $m_k$-fold tensor product with weights
$\L_{j(1)}, ..., \L_{j(m_k )} $ which must satisfy

$$e^{-2\pi i \l_i \cdot \l_k } = e^{-2\pi i \l_i \cdot \sum_{p=1}^{m_k}
\l_{j(p)}}.     \eqno(3.2) $$

\ni An example at level 2 was furnished by (6.8) of \OTUb but the detailed
construction of the state will, fortunately, be immaterial for the results
that follow. We label  the spaces 1, 2, ...,
$m_k$, respectively. So

$$\fih  = \fih_{(1)} + \fih_{(2)} +  \cdots  + \fih_{(m_k)}, \eqno(3.3) $$

\ni and similarly for the Heisenberg subalgebra,

$$\EM = \hat E_{M(1)} + \hat E_{M(2)} +  \cdots  + \hat E_{M(m_k)}.
\eqno(3.4)$$

In this notation the bracketed subscript denotes which of the spaces
it is upon which the operator acts non-trivially. Thus the terms in (3.3)
mutually commute and each have vanishing square as each is given by the
vertex operator construction (3.1). Hence

$$\eqalign{{(\fih)^{m_k} \over m_k ! } & =  \fih_{(1)} \fih_{(2)}
\ldots\fih_{(m_k)}\cr
& =  e^{-2\pi i \l_i \cdot \l_k }  Y^i_{(1)} \ldots Y^i_{(m_k)} Z^i_{(1)}
\ldots Z^i_{(m_k)}\cr } \eqno(3.5) $$

\ni using (3.2). Finally, using (3.1b) and (3.4),

$$ {(\fih)^{m_k} \over m_k !} = e^{-2\pi i \l_i \cdot \l_k } Y^i Z^i.
\eqno(3.6) $$

\ni This is the new vertex operator construction announced earlier. Of course
 (3.1a) is now seen to be the special case of (3.6) when $m_k = 1$. More
generally, given an  irreducible representation with highest weight $\L$, at
level $x$,

$$  {(\fih)^{x} \over x!} = e^{-2\pi i \l_i \cdot \l } Y^i Z^i. \eqno(3.7) $$

In the numerator of the single soliton solution we see that
 the coefficient of the highest
power of $QW_i $ is given by the expectation value of  (3.6) with respect to
the weight state $\rlk $. Since the state is annihilated by all the step
operators for positive roots of $\gh$, that is, by all the generators of
positive principal grade, it is certainly  annihilated by those elements of
the Heisenberg subalgebra (3.4) with positive suffix. Hence $Y^i$ yields
unity, and

$$ {\llk (\fih)^{m_k} \rlk  \over m_k ! } = e^{-2\pi i \l_i \cdot \l_k } ,
\eqno(3.8) $$

\ni a surprisingly simple result, holding for all $k=0,1,2,\dots r$ and
$i=1,2,\dots r$. Noting that (3.8) is independent of $z$ (or $\eta$). This is
a special case of a more general result: the expectation value of any power
of $\fih$ is independent of $\eta$. As we noted, the operator $d'$, (2.1),
plays the role of the Lorentz boost for solitons. More explicitly, from the
definitions,

$$ [d' , \fih ] = {d \over d\eta } \fih. $$

\ni Hence

$$ [d' , (\fih)^a  ] =  {d \over d\eta } (\fih)^a.$$

\ni The expectation value of this vanishes as $d'$ has equal, real
eigenvalues to each side. Thus $\llk (\fih)^a \rlk $ is independent of the
rapidity $\eta$ and so is Lorentz invariant. Using (3.8), the solution for
the single soliton created by $g(0)=expQ\fih$ in (2.10) has the form

$$e^{-\b \l_k \cdot \Phi } = {1 +  \cdots  +
e^{-2\pi i \l_i \cdot \l_k }(QW_i)^{m_k}
\over
                              \left(1  + (QW_i)\right)^{m_k}    }
      \eqno(3.9)$$

\ni where the dots indicates powers of $QW_i $ intermediate between 0 and
$m_k$. These intermediate  powers, whose coefficients we have not
calculated, do not affect the asymptotic limit $x \rightarrow \pm \infty $,
which are equivalent to $W_i  \rightarrow \infty $ or $W_i \rightarrow 0$.
Thus

$$ e^{-\b \l_k \cdot \Phi} =
\cases{ e^{-2\pi i \l_i \cdot \l_k } &$x \rightarrow \infty$    \cr
                                   1 &$x \rightarrow -\infty$   \cr} $$

\ni In particular, this shows that,  asymptotically, $\Phi $ does approach
one of  the \lq\lq degenerate vacua" , so satisfying the boundary conditions
expected of a soliton solution.

This result can readily be extended to solutions with any number of solitons.
The operator product of two vertex operators (3.6) is

$$ {(\fih)^{m_k} \over  m_k !}{(\hat F^{i'}(\h'))^{m_k}\over m_k !} = [ X_{i
, i'}(z,z') ]^{m_k}
     :{(\fih)^{m_k} \over  m_k !}{(\hat F^{i'}(\h'))^{m_k}\over m_k !}: $$

\ni where $|z| > |z'|$ i.e. $\eta ' > \eta $, where the notation of section
5 of \OTUb
 for the complex numbers $X_{ij}$ is
used.

Hence the two soliton solution created by $\fih $ and $\hat{F}^{i'} (\eta ')$
takes the form

$$e^{-\b \l_k \cdot \Phi } = {1 +  \cdots  + e^{-2\pi i (\l_i  +
\l_{i'})\cdot \l_k } (QW_i  Q'W_{i'} )^{m_k} X_{i,i'}(z,z')^{m_k} \over\left(
  1 +  QW_i+Q'W_{i'}  + QW_i  Q'W_{i'}  X_{i,i'}(z,z')\right)^{m_k}  }     $$

\ni with asymptotic limits $\exp[-2\pi i ( \l_i + \l_{i'} )\cdot\l_k ] $ and
1. Again this confirms that the solution interpolates
degenerate vacua. This argument is readily extended to more solitons
using the  work of section 5 of \OTUb.
\bigskip

\def\yih{Y^i(\eta)}
\def\zih{Z^i(\eta)}
\def\fibh{\hat F^{\bar \i}(\eta + i\pi)}
\def\yibh{Y^{\bar \i}(\eta + i\pi)}
\def\rlj{|\Lambda_j>}\def\rlk{|\Lambda_k>}
\def\llj{<\Lambda_j|}\def\llk{<\Lambda_k|}
\def\rlo{|\Lambda_0>}
\def\llo{<\Lambda_0|}

\bigskip
\ni{\it 4. The Crossing Relations for Solitons}
\medskip

This identity will arise naturally in extending the result (3.7) to lower
powers of  $\fih $. Consider a representation of level $x$ with highest weight
$\L$, constructed again as occurring in the decomposition of tensor product
of $x$ fundamental representations of level 1. Consider a power $a$ of $\fih$
less then $x$: $$
{(\fih)^a \over a! } = \fih_{(1)} \fih_{(2)} \cdots \fih_{(a)} + \cdots
, \eqno(4.1) $$

\ni where the dots indicates extra terms of a similar structure obtained by
assigning the ``$a$'' quantities $\fih$ separately to the $x$ different
species. There are consequently $ \pmatrix{ x \cr a\cr} $ terms on the right
hand of (4.1) in all. Similarly, if we consider the power $x-a$:
$$
{(\fih)^{x - a} \over (x - a )! } = \fih_{(a+1)} \fih_{(a+2)} \cdots
\fih_{(x)} + \cdots , \eqno(4.2) $$

\ni The first terms on the right hand side of (4.1) and (4.2) multiplied give
$(\fih)^x / x!$ are thus naturally complementary. There is a similar  pairing
between any term on the right hand side of equation (4.1) and one on the
right hand side of (4.2). Thus the respective number of terms $ \pmatrix{ x
\cr a\cr} $ and  $ \pmatrix{ x \cr x - a\cr} $ must be equal as indeed they
are. We shall now relate the first two terms exhibited above so that a similar
relation will follow for all the $ \pmatrix{ x \cr  a\cr} $ complementary
pairs
$$\displaylines{
\fih_{(1)} \fih_{(2)} \cdots \fih_{(a)}
   = e^{-2\pi i\l_i \cdot \left(\sum_{p=1}^a \l_{j(p)}\right)}
         Y^i_1(\eta) \cdots  Y^i_a(\eta)Z^i_1(\eta) \cdots Z^i_a (\eta)
\hfill\cr =e^{-2\pi i\l_i \l} Y^i(\eta )\Bigl\{ e^{2\pi i\l_i \cdot
\left(\sum_{p=a+1}^x \l_{j(p)} \right)}
       Y^i_{a+1}(\eta)^{-1}\cdots  Y^i_x(\eta)^{-1} Z^i_{a+1}(\eta)^{-1}
\cdots Z^i_x(\eta)^{-1}  \Bigr\} Z^i(\eta) .\cr
\hfill (4.3)\cr}$$

\ni Notice that the factors outside the curly bracket are independent of the
subset of $x$ chosen since $\yih $ and $\zih $ are expressed by
exponentiating (3.4). We shall relate the terms inside the curly brackets
to the first term on the right hand side of (4.2) for a suitable new choice
of species and rapidity.

Consider a typical factor in the curly brackets in (4.3) associated with one
of the species whose tensor product is taken, with label $p$, say, namely
$$e^{2\pi i\l_i\cdot\l_{j(p)}}(Y^i(\eta)_{(p)})^{-1}(Z^i(\eta)_{(p)})^{-1}.$$
 We shall temporarily drop the label $p$ and verify the following remarkable
identity for this factor:
$$
e^{2 \pi i\l_i \cdot \l_j} (\yih)^{-1} (\zih)^{-1} = \fibh . \eqno(4.4)  $$

\ni Thus,  the contents of the curly brackets in
(4.3) can be written
$$
\hat F_{(a+1)}^{\bar \i}(\eta +i\pi)
\hat F_{(a+2)}^{\bar \i}(\eta +i\pi)  \cdots
\hat F_{(x)}^{\bar \i}(\eta +i\pi).
\eqno(4.5)
$$

To prove (4.4) note that, by (3.1),
$$
(\yih)^{-1} = \exp -\sum_{N>0} {\g_i \cdot q([N]) z^N \hat E_{-N} \over N} ,
\eqno(4.6a)$$

\ni where, as in (2.8),
$$
z = i e^{\eta}e^{ -{i\pi(1 + c(i)) \over 2h}} .\eqno(4.6b) $$

\ni Now, $-\g_i $ lies on the orbit of $\s$ whose representative element is
defined as $\g_{\bar \i}$. As observed in the study of the quantum particle
excitations, $\bar \i$ is interpreted as the antispecies of $i$ \FLO.
Precisely
\ref\FO{ A. Fring and D.I. Olive,  {\it Nucl. Phys.} {\bf B379}
(1992) 429-447, \lq\lq The Fusing Rule and the Scattering Matrix of Affine
Toda Theory"}
  $$
-\g_i = \s^{{h \over 2} - {c(i) - c(\bar \i) \over 4}} \g_{\bar \i}.
\eqno(4.7) $$

\ni Recalling (2.4) we see that
$$
\yih^{-1} = \exp \sum_{n>0}
  {\g_{\bi}\cdot q([N])
     \left[ z e^{{-2\pi i \over h}
           \left({h \over 2} - {c(i) - c(\bi  ) \over 4} \right) }
     \right]^N \hat E_{-N}\over N }.
 $$

\ni Now, by (4.6b), the argument of the square brackets is equal to
$$
i e^{\eta} e^{-{i\pi \left(1+c(i)\right) \over 2h}}
e^{-\pi i} e^{{i\pi \left(c(i) - c(\bi)\right) \over 2h}} =
i e^{\eta +i\pi} e^{-{i\pi \left(1+c(\bi)\right) \over 2h}} = -
\bar z.
$$

\ni  where $\bar z$ differs from $z$ defined by (4.6b) only in the species
$\bi$ replacing its antispecies $i$. Hence, $\yih^{-1} = \yibh $ and
similarly for $\zih^{-1}$. Finally,  in order to prove
(4.4) we need to show $$
e^{2\pi i \l_i \cdot \l_j} = e^{-2\pi i \l_{\bi} \cdot \l_j} \eqno(4.8)
$$

\ni and this relies on the fact that the map $i \rightarrow \bi $ is a
symmetry $\tau$, say, of the ordinary Dynkin diagram of $\gg$. Hence this can
be
lifted to a linear automorphism of the root system, also called $\tau$, which
is
either outer or equal to the identity, such that $\a_{\bi} = \tau(\a_i)$.
Comparing with (4.7), it follows that $\tau = -\s_0$, where $\s_0$ can be
calculated but is known as that special element of the Weyl group of $\gg$
which maps the positive Weyl chamber into its negative. As a consequence,
$$
\l_{\bi} = \tau(\l_i) = - \s_0(\l_i).$$

\ni (4.8) follows from this as $\s_0(\l_i)$ differ from $\l_i$ by an element
of the root lattice. Thus, (4.4) is proven and, using (4.3) and (4.5), we
conclude
$$
{(\fih)^a \over a!} = e^{-2\pi i \l_i \cdot \l} \yih {(\fibh)^{x-a} \over
(x-a)!} \zih . \eqno(4.9)
$$

If we take expectation values of (4.9) with respect to the highest weight
state $|\L>$, the dependence on the rapidity drops out as we explained,
leaving
$$
{<\L| (F^i )^a |\L> \over a!} =  e^{-2\pi i \l_i \cdot \l} {<\L|
(F^{\bi})^{x-a} |\L> \over (x-a)!} . \eqno(4.10)
$$

\ni This relates coefficients in the single soliton solution. As  it relates
high powers of $\fih$ to low powers, it simplifies calculations of these
solutions.

Since $\yih$ and $\zih$ appearing in (4.9) are independent of the integer
$a$, we can deduce the following relation between group elements:
$$
e^{Q\fih} = e^{-2\pi \l_i \cdot \l} Q^x \yih e^{{1 \over Q} \fibh} \zih.
\eqno(4.11)
$$

\ni This is our main result, being the surprising, nonlinear relation between
$\fih$ and $\fibh$. We shall argue that it expresses the ``crossing
property'' of a soliton solution into an antisoliton. This is clearer if we
rewrite it as
$$
e^{Q\hat F^{\bi}(\eta)} = e^{-2\pi \l_{\bi} \cdot \l} Q^x Y^{\bar \i}(\eta)
e^{{1 \over Q} \hat F^i(\eta + i \pi)} Z^{\bi}(\eta) .\eqno(4.12)
$$

This relates the analytic continuation in rapidity, $\eta \rightarrow \eta
+i\pi$, of the group element $e^{{1 \over Q} \fih}$ creating a soliton of
species $i$, to the group element $e^{Q \hat F^{\bi}(\eta)}$ creating an
antisoliton of species $\bi$. The remaining factors $Y^{\bi}(\eta) ,
Z^{\bi}(\eta)$ can be thought of as ``crossing matrices''. The reason for
saying this is that in a local quantum field theory, the S-matrix
describing the scattering of particles obeys a ``crossing property'', more
precisely \lq\lq the substitution rule for crossed processes" \ELOP. In two
dimensions, the S-matrix for the elastic  scattering
of an antiparticle on a target is obtained by the analytic continuation $\eta
\rightarrow \eta + i\pi$ of the S-matrix for the elastic scattering of the
particle on the same target. If the particle has internal degrees of freedom,
then
an additional crossing matrix multiplies the S-matrix.

The equation (4.11) is obviously a very similar relation. One difference is
that the crossing property just described, holds in the quantum theory as
the S-matrix elements are probability amplitudes. The solitons are the only
sort of particles possible in a classical theory such as we consider. Thus,
(4.12) is actually the classical analogue of the crossing property and is
remarkable because it is formulated purely in terms of the underlying
Kac-Moody group. What is attractive about this is that, hitherto, the
crossing property of the affine Toda S-matrix has been incorporated in a
rather ad-hoc way. Relation (4.12) suggests the possibility of a more
systematic understanding.

We now justify our interpretation by using the crossing identity (4.12)
inside the expression for the multisoliton solution in order to see
explicitly how analytic continuation through $i\pi$ in the rapidity of one
soliton of species $i$ yields the multisoliton solution in which the
soliton of species $i$ is replaced by its antisoliton with species $\bi$.
The rapidities are unchanged but the complex numbers $Q_1$, $Q_2$, $\ldots$
associated with each soliton undergo a transformation which will be
determined.

Consider the $n$ soliton solution created by the group element
$$
g(0) = e^{Q_1 \hat F^{i_1}(\eta_1)} \cdots
e^{Q_{k-1} \hat F^{i_{k-1}}(\eta_{k-1})}
e^{Q \hat F^{\bi}(\eta)}
e^{Q_{k+1} \hat F^{i_{k+1}}(\eta_{k+1})} \cdots
e^{Q_n \hat F^{i_n}(\eta_n)}.
$$

\ni The $k^{th}$ factor creates the antisoliton of species $i$ to be
crossed into an soliton of species $i$ and is labelled in the simple way
indicated. Inserting the crossing identity (4.12)
$$
g(0) = e^{-2\pi i \l_{\bi} \cdot \l} Q^x Y^{\bi}(\eta) \tilde g(0)
Z^{\bi}(\eta)$$

\ni where
$$\tilde g(0) = e^{\tilde Q_1 \hat F^{i_1}(\eta_1)} \cdots
e^{\tilde Q_{k-1} \hat F^{i_{k-1}}(\eta_{k-1})}
e^{\tilde Q \hat F^i(\eta + i\pi)}
e^{\tilde Q_{k+1} \hat F^{i_{k+1}}(\eta_{k+1})} \cdots
e^{\tilde Q_n \hat F^{i_n}(\eta_n)}.
$$
is similar to $g(0)$ but with the $k$'th factor now of the form $e^{\tilde Q
\hat F^i(\eta + i\pi)}$ , being the analytic continuation of the factor for
the creation of a soliton of species $i$. The quantities $Q$ have all
changed: $$ \cases{\tilde Q = 1/Q  & \cr
      \tilde Q_r = Q_r  X_{i_r \bi }  &, $r\neq k$. \cr} \eqno(4.13)
$$

\ni We have used the normal ordering relations
$$
\eqalign{Z^{\bi}(z) F^j(z_j) &= F^j(z_j) Z^{\bi}(z) X_{{\bi} j } (z, z_j)\cr
F^j(z_j) Y^{\bi}(z) &=  Y^{\bi}(z) F^j(z_j) X_{{\bi} j } (z, z_j) \cr}
$$

\ni where $X_{ij}(z,\zeta) $ is defined in section 5 of \OTUb. Finally we use
$$
e^{-\mu \e1 x^+ }Y^{\bi}(\eta) \tilde g(0) Z^{\bi}(\eta) e^{-\mu \em1 x^- } =
(W_{\bi}(\eta))^{-x}Y^{\bi}(\eta)e^{-\mu \e1 x^+ } \tilde g(0)
e^{-\mu \em1 x^- }
Z^{\bi}(\eta) $$

\ni which follows from (2.2) and the definitions. Putting all this
together
$$
\displaylines{\quad<\L_k| e^{-\mu \e1 x^+} g(0) e^{-\mu \em1 x^-}|\L_k>
\hfill\cr
 \hfill{}= e^{-2\pi i\l_{\bi} \cdot \l_k } \left({Q \over W_{\bi}}
\right)^{m_k} <\L_k| e^{-\mu \e1 x^+ } \tilde g(0) e^{-\mu \em1 x^-
}|\L_k>.\quad\cr} $$

\ni When we insert this into (2.10), the factors $(Q/W_{\bi})^x$ cancel
leaving  the solition  created by
$g(0)$ equalling that created by $\tilde g(0)$, apart from a factor
$e^{-2\pi i\l_{\bi} \cdot \l_k }$
which merely indicates that the vacua are interpolated in reverse order. This
then demonstrates that a soliton solution crosses into another solution with
one soliton replaced by the antispecies under the analytic continuation $\eta
\rightarrow \eta +i\pi$. This procedure is accompanied by the transformation
(4.13) on the $Q$'s. Since they describe not only the coordinate of the
soliton, but also its internal degrees of freedom, this is to be viewed as
the effect of the crossing matrix. It is intriguing that for the soliton
which is crossed, $Q \rightarrow 1/Q$ which reminiscent of a modular
transformation.

Because relativistically invariant soliton theories play a prefered  r\^ole
within their integrable hierarchies according to the ideas of
Zamolodchikov \ZamC, and because the concept of antiparticles and crossing
are special to relativistic theories, one might anticipate interesting
structure to emerge. What we have shown is that in affine Toda theories the
crossing relation is embodied in the remarkable identity (4.12) which is
purely a statement about the underlying affine Kac-Moody algebra and a
highest weight representation of arbitrary level. As we already know  that
the ``fusing" rules for trilinear couplings are embodied in the operator
product expansion of the $\fih$, \OTUb, it would  appear that, for affine
Toda solitons, there is the possibility of a purely group theoretical
understanding of the   principles of S-matrix theory. \medskip
\noindent{\bf Acknowledgements}
\medskip
Marco Kneipp is grateful to CNPq (Brazil) for financial support. We wish
to thank Peter Johnson, Neil Turok and Jonathan Underwood for discussions.
\medskip
\listrefs
\bye